\documentclass[]{article}
\usepackage{emulateapj}
\usepackage{epsfig}
\usepackage{ifthen}

\submitted{Accepted by The Astrophysical Journal}

\def\gsim{\;\rlap{\lower 2.5pt
 \hbox{$\sim$}}\raise 1.5pt\hbox{$>$}\;}
\def\lsim{\;\rlap{\lower 2.5pt
   \hbox{$\sim$}}\raise 1.5pt\hbox{$<$}\;}
\def\msol{{\rm\,M_\odot}}

\def\kpc{{\rm\,kpc}}

\def\spose#1{\hbox to 0pt{#1\hss}}
\def\lta{\mathrel{\spose{\lower 3pt\hbox{$\mathchar''218$}}
     \raise 2.0pt\hbox{$\mathchar''13C$}}}
\def\gta{\mathrel{\spose{\lower 3pt\hbox{$\mathchar''218$}}
     \raise 2.0pt\hbox{$\mathchar''13E$}}}

\begin{document}
	
\title{Problems for MOND in Clusters and the Ly$\alpha$ Forest ?}

\author{Anthony Aguirre, Joop Schaye \& Eliot Quataert\altaffilmark{1}}
\affil{School of Natural Sciences, Institute for Advanced Study, Princeton, NJ 08540\\
aguirre@ias.edu, schaye@ias.edu, eliot@ias.edu}
\altaffiltext{1}{Chandra Fellow}
  
\setcounter{footnote}{0}

\begin{abstract}

	The observed dynamics of gas and stars on galactic and larger
scales cannot be accounted for by self-gravity, indicating that there
are large quantities of unseen matter, or that gravity is
non-Newtonian in these regimes.  Milgrom's MOdified Newtonian Dynamics
(MOND) postulates that Newton's laws are modified at very low
acceleration, and can account for the rotation curves of galaxies and
some other astrophysical observations, without dark matter.  Here we
apply MOND to two independent physical systems: Ly$\alpha$ absorbers
and galaxy clusters.  While physically distinct, both are simple
hydrodynamical systems with characteristic accelerations in the MOND
regime.  We find that, because MOND violates the strong equivalance
principle, the properties of Ly$\alpha$ absorbers depend strongly upon
the (unknown) background acceleration field in which they are
embedded.  If this field is small compared to their internal
accelerations, then the absorbers are more dense and about ten times
smaller than in Newtonian gravity with dark matter, in conflict with
sizes inferred from quasar pair studies.  If, however, the background
field is rather large, then the absorbers take on properties similar
to those predicted in the CDM picture.  In clusters MOND appears to
explain the observed (baryonic) mass-temperature relation.  However,
given observed gas density and enclosed mass profiles and the
assumption of hydrostatic equilibrium, MOND predicts radial
temperature profiles which disagree badly with observations.  We show
this explicitly for the Virgo, Abell 2199 and Coma clusters, but the
results are general, and seem very difficult to avoid.  If this
discrepancy is to be resolved by positing additional (presumably
baryonic) dark matter, then this dark matter must have $\sim1-3$ times
the cluster gas mass within $1\,$Mpc, and about ten times the gas mass
with $200\,$kpc.  This result strongly disfavors MOND as an
alternative to dark matter.

\end{abstract}

\section{Introduction}

	The currently most widely-accepted `standard model' of
cosmology holds that the vast majority of the mass density of the
universe is hidden in dark forms.  The rotation curves of galaxies and
the dynamics of galaxy clusters cannot be accounted for by the
gravitation of visible stars and gas, while constraints from
primordial nucleosynthesis studies imply that the additional `dark
matter' postulated to remedy this discrepancy must be non-baryonic.
On a cosmological level, collisionless (or very weakly collisional)
dark matter is required if primordial density perturbations of
amplitude $\Delta\rho/\rho \sim 10^{-5}$ are to grow quickly enough to
form galaxies by the present epoch.  Finally, recent determinations of
the high redshift type Ia supernova Hubble diagram, in tandem with
microwave background data implying a flat cosmic geometry, and a large
collection of data indicating that clustering matter only contributes
$\sim 30\%$ of the critical density, imply that the universe also
contains `dark energy' of a yet more exotic form which causes
acceleration in the cosmic expansion (see, e.g., Peebles 1999 and
Turner 1999 for recent reviews).

	Discomfort with this repeated postulation of invisible matter
with increasingly unusual properties has led some, quite reasonably,
to ask whether the observed phenomena could be accounted for not by
the presence of unseen matter, but by a departure from
Newtonian/Einsteinian dynamics in the regime where dark matter is
hypothesized to be important.  Perhaps the most successful of such
proposals is the MOdified Newtonian Dynamics (MOND) proposed by
Milgrom (1983a,b,c), in which Newtonian dynamics breaks down below an
acceleration threshold of $a_0 \sim 10^{-8}{\,\rm cm\,s^{-2}}$.
Successes of this theory are that it accounts for the rotation curves
of galaxies of various luminosities (Sanders \& Verheijen 1998) and
surface brightnesses (de Blok \& McGaugh 1998; McGaugh \& de Blok
1998), accounts for the Tully-Fisher and Faber-Jackson relations (van
den Bosch \& Dalcanton 2000; Sanders 1996, 2000), and -- arguably --
roughly accounts for the amount of dark matter inferred in clusters
(Sanders 1999).

	Despite some attempts, MOND has not been generalized into a
satisfactory relativistic theory which can yield unambiguous
cosmological predictions.  Scott et al. (2001) explore the
difficulties in applying MOND to cosmology and review a number of
claimed conceptual and empirical difficulties with MOND.  But in light
of MOND's general success when applied to galaxies, and the current
lack of any decisive empirical argument against it, it is worth
investigating whether MOND works in detail in systems for which it was
not designed yet makes relatively unambiguous predictions, and for
which good observations are available.  We propose and perform two
such tests.  First, we deduce basic properties of the Ly$\alpha$
absorbers, making use of the technique of Schaye (2001).  These can be
compared directly to observations concerning the sizes and number
densities of the absorbers.  Second, we derive relations between the
density, enclosed mass, and temperature profiles of galaxy clusters
which can be directly compared to available X-ray data.  Both
Ly$\alpha$ absorbers and clusters constitute relatively simple
physical systems, are well within the MOND regime, and (in the absence
of dark matter) are dominated by gas which can be accurately observed,
making them ideal testing grounds for MOND.

\section{MOND and the Ly$\alpha$ Forest}

As argued by Schaye (2001; hereafter S01), basic properties of the gas
responsible for Ly$\alpha$ absorption in quasar spectra can be deduced
using simple physical arguments. Along any sightline passing
through a gas `cloud', the size of the region dominating the absorption
will typically be of order the local Jeans length, regardless of the
overall shape of the cloud and regardless of whether the cloud as a
whole is in dynamical equilibrium. Using this reasoning, S01
calculates properties of the Ly$\alpha$ absorbers which are in good
agreement with numerical simulations and available observations.  Because
the absorbers are dynamically simple and have very low characteristic
accelerations, MOND makes strong predictions about their properties, which
we will now derive in parallel to the treatment by S01.
          
MOND can be formulated in a number of ways, either as a modification
of inertia (e.g., Milgrom 1999) or of gravity (e.g., Bekenstein \&
Milgrom 1984).  In its most general incarnation, the gravitational
acceleration in an `isolated' (meaning not embedded in a gravitational
field with larger characteristic acceleration) system is given by
\begin{equation}
a=\sqrt{a_Na_0}
\label{eq-monda}
\end{equation}
when $a_N \ll a_0$, where $a_N$ is the acceleration calculated using
Newtonian gravity, and $a_0\approx 1.2\times 10^{-8}\,{\rm
cm\,s^{-2}}$ (McGaugh \& de Blok 1998) is the MOND acceleration
parameter. (Systems which {\em are} embedded in a large external field
are discussed below).  This acceleration law yields a dynamical time
\begin{equation}
t_{\rm dyn} \equiv \left({L\over a}\right)^{1/2}= \left({L \over \rho a_0 G}\right)^{1/4} = 
	\left[{L(1-Y)\over n_Hm_pa_0G}\right]^{1/4}, 
\label{eq-tdyn}
\end{equation}
where $L$ and $\rho$ are the characteristic size and density of the
system, respectively, $n_H$ is the hydrogen number density in a medium
of hydrogen mass fraction $(1-Y)$, and $m_p$ is the proton mass. 
The system's sound-crossing time is unchanged by MOND and is
\begin{equation}
t_{\rm sc} \equiv L/c_s = L\left({\mu m_p\over \gamma k T}\right)^{1/2},
\end{equation}
where $T$ is the temperature, $\gamma = 5/3$ is the assumed adiabatic
index and $\mu \approx 0.59$ is the mean molecular weight per particle
for a fully ionized primordial plasma with $Y=0.24$. Setting these
timescales equal to each other yields the Jean length
\begin{equation}
L_J=\left({1-Y\over n_H a_0 G}\right)^{1/3}\left({\gamma k T\over\mu}\right)^{2/3}m_p^{-1}.
\label{eq-mondjl}
\end{equation}
This can be converted to a column density $N^J_{H I}=L_J n_H \times (n_{H
I}/n_H)$ using the ionization correction from S01,
valid for a highly ionized, optically thin plasma:
\begin{equation}
{n_{H
I}\over n_H}=n_H {1-Y/2 \over 1-Y}{\beta \over \Gamma},
\end{equation}
where $\beta \approx 4\times10^{-13}T_4^{-0.76}{\rm cm^3\,s^{-1}}$ and
$\Gamma \equiv \Gamma_{\rm 12}\times 10^{-12}{\,\rm s}$ are the
recombination and ionization rates with $T_4\equiv T/10^4\,$K;
$\Gamma_{\rm 12} \approx 1$ is measured at redshift $z \sim 3$ (see Scott et
al. 2000 and references therein). The resulting Jeans column density
can be expressed as a function of overdensity $\delta \equiv n_H/\bar
n_H-1$, as
\begin{eqnarray}
\nonumber
N_{H I}^J &=& \left({\gamma k T\over a_0^{1/2}G^{1/2}\mu m_p^4}\right)^{2/3}
(1-Y/2)(1-Y)\left({\beta\over \Gamma}\right) \\ \nonumber
&& \times [\rho_c^0\Omega_b(1+\delta)]^{5/3}(1+z)^5 \\ \nonumber
&\approx& 3.6\times10^{12}{\,\rm cm^{-2}}\,T_4^{-0.09}\Gamma_{12}^{-1}\left({\Omega_b h^2\over 0.02}\right)^{5/3} \\
&& \times(1+\delta)^{5/3}\left({1+z\over 4}\right)^5,
\end{eqnarray}
where $T_4\equiv T/10^4\,$K, $\rho_c^0$ is the current critical
density and $\Omega_b$ is the baryonic density parameter. Thus
isolated Ly$\alpha$ absorbers of density equal to the cosmic mean have about
ten times lower column density than in the CDM picture, with different
dependences on $T$, $z$, etc. (c.f. S01, Eq. 10).  One can also
express the Jeans length in terms of the observed H\,I column density:
\begin{eqnarray}
\nonumber
L_J&=&\left[{(1-Y)(1-Y/2)\over a_0^2G^2m_p^6}\left({\gamma k T\over \mu}\right)^4\left({\beta\over \Gamma}\right)N_{H I}^{-1}\right]^{1/5} \\ 
&\approx& 11\,{\rm kpc}\left({N_{H I}\over10^{14}{\rm cm^{-2}}}\right)^{-1/5}
\Gamma_{12}^{-1/5}T_4^{0.65},
\label{eq-absz}
\end{eqnarray}
about a factor of ten smaller than an absorber of the same column
density in the CDM picture (S01, Eq. 12), and with different scalings
for all parameters.  Note that significant external pressure would
only decrease this size.  A few self-consistency checks are in order.
First, the Newtonian acceleration is
\begin{equation}
a_N \approx 4\times10^{-13}{\rm cm\,s^{-2}}\left({N_{H I}/10^{14}{\rm cm^{-2}}}\right)^{2/5}T_4^{0.70}\Gamma_{12}^{2/5},
\label{eq-aint}
\end{equation}
verifying that the system is in the MOND regime.  The internal acceleration is then
$\approx a_0/170$.  Second, the system's dynamical time is
\begin{equation}
t_{\rm dyn} = 2\times 10^{16}\,{\rm s}\left({N_{H I}/10^{14}{\rm cm^{-2}}}\right)^{-1/5}
T_4^{0.15}\Gamma_{12}^{-1/5},
\end{equation}
whereas the Hubble time\footnote{We assume that whatever cosmology
MOND engenders will be a Friedmann model with a scale factor that
evolves roughly as in a standard cosmology with
$\Omega_m=1-\Omega_\Lambda=0.3$ (as indicated by observations).}  is
$\approx (1.1-2.7)h_{65}^{-1}\times 10^{17}\,$s for $3 \ga z \ga 1$,
so the absorbers are self-consistently in local hydrostatic equilibrium
(assuming they can {\em reach} this equilibrium; see below).

The method of S01 can also be used to
`invert' the observed column density distribution 
into an estimate of the total mass density in Ly$\alpha$ absorbers, using
\begin{eqnarray}
\nonumber
\Omega_{\rm gas}&=&{8\pi G H(z)\over 3 H_0^2 c (1+z)^2}\left({
\gamma k T m_p \Gamma \over \mu(1-Y)(1-Y/2)\beta}\right)^{2/5}
\\
&&\times(a_0G)^{-1/5}\int dN_{HI} N_{HI}^{2/5}{d^2n(N_{HI},z)\over dN_{HI}dz}
\label{eq-omgas}
\end{eqnarray}
where $d^2n(N_{HI},z)/dN_{HI}dz$ is the differential number of
observed lines of column density $N_{HI}$ at redshift $z$.  

	The preceding calculation applies to an isolated absorber, but
because the strong equivalence principle is violated in MOND, the internal
dynamics of systems can change if they are embedded in an external
acceleration field $\vec g_0$, even if it is homogeneous (i.e., $\vec
g_0 = g_0\hat z$ with coordinates such that $\hat z$ is the external
field direction).  Milgrom (1986) shows that in the Lagrangian
formulation of MOND (Bekenstein \& Milgrom 1984), if $g_0$ is small
compared to $a_0$ but large compared to the internal accelerations of
a system, then gravity in the subsystem is Newtonian in
the coordinate system $\{x',y',z'\}\equiv\{x,y,z/2\}$ with an
effective Newton's constant $G'= (a_0/g_0) G$.  In this case,
properties of the absorbers in MOND can be computed directly from the
corresponding Newtonian/CDM expressions from S01, with those
substitutions.\footnote{This sort of behavior is not unique to the
Lagrangian formulation; a similar result can be derived from the simple
formulation of MOND given by Eq.~\ref{eq-monda}.}  For example, the
size of the absorbing region is then
\begin{equation}
L_J \approx 40{\rm\,kpc}\left({N_{H I}\over10^{14}{\rm cm^{-2}}}\right)^{-1/3}
\left({a_0/g_0\over 25}\right)^{-2/3}\Gamma_{12}^{-1/3}T_4^{0.41}
\label{eq-absz2}
\end{equation}
in the direction perpendicular to the external field, and somewhat
(about two times) longer in the parallel direction.  The density
parameter in Ly$\alpha$ absorption systems is given by the CDM
calculation of S01, adjusted by a factor of
$(a_0/g_0)^{-1/3}f_g^{-1/3}$, where $f_g\approx 0.16$ is the
gas-to-matter mass ratio in the CDM calculation.

	The above analysis gives two quite testable predictions
concerning the Ly$\alpha$ forest in MOND.  First, the total density of
gas can be computed using Eq.~\ref{eq-omgas} (for the isolated case)
or Eq. 16 of S01 (for the external field case).  Since Ly$\alpha$
systems are very deep in the MOND regime, there is no {\em a priori}
reason to expect that this density will be at all reasonable (i.e., as
compared to the nucleosynthesis value of $\Omega_b$), and this can be
assessed.  Second, the characteristic sizes of absorbers, given by
Eqs.~\ref{eq-absz} and~\ref{eq-absz2}, can be compared to observations
of lensed quasars and quasar pairs that constrain the transverse sizes
of absorbers.  In making both comparisons, the external field case
requires a value for the mean acceleration field of Ly$\alpha$
absorbers at $z=3.$ This is currently not calculable, but may be
crudely estimated.  At $z=0$, typical mean observed accelerations can
be obtained by dividing typical bulk flow velocities of $\sim
600\,{\rm km\,s^{-1}}$ (see, e.g., Dekel et al. 1999; Dale et
al. 1999) by a Hubble time, yielding $g_0/a_0 \approx 10^{-2}$; we
shall take an upper limit of $g_0/a_0 \la 1/50$.  If the fluctuations
in the Newtonian gravitational potential (which in linear theory are
constant in time in an Einstein--de Sitter cosmology) do not shrink,
then $g_0(z)/a_0 \la 10^{-2}(1+z)^{1/2}$.  This is an upper limit, and
the accelerations could be significantly smaller at $z=3$ in a MOND
universe if the potential fluctuations have grown considerably since
then.  From Eq.~\ref{eq-aint} above, we see that an external field
$g_0 \ga a_0/170 $ is required to modify the absorber internal
dynamics.  Thus we can consistently consider external fields with $25
\la a_0/g_0 \la 170.$

We have computed the total gas density in Ly$\alpha$ absorbers in both
the isolated and non-isolated cases using the data of Hu et al. (1995)
and Petitjean et al.  (1993).  In the isolated case we find
$\Omega_{\rm gas} \approx 0.008-0.009$. This is somewhat smaller than
the $\Omega_{\rm gas} \approx 0.045$ inferred from the same data in
the CDM picture or using nucleosynthesis, but is still
plausible -- even if we assume that most baryons must be in gas in
the IGM -- considering the uncertainty inherent in the
analysis and the neglect of underdense and collisionally ionized hot
gas.  In the case of a significant external field, we find
\begin{equation}
\Omega_{\rm gas} \approx 0.028\left({a_0/g_0\over
25}\right)^{-1/3}T_4^{0.59}\Gamma_{12}^{1/3}h_{65}^{-2},
\end{equation}
giving $0.005 \la \Omega_{\rm gas} \la 0.03$; the higher end is in
comfortable agreement with observational constraints.
	
	For isolated absorbers, MOND also predicts that the low column
density ($\sim 10^{14}\,{\rm cm^{-2}}$) absorbers have a rather small
characteristic size: $\sim 10\kpc$ versus $\sim 100\kpc$ for Newtonian
gravity with CDM.  We have argued that in a plausible external field,
the absorbers should have sizes of $20-80\,$kpc in their long
direction, and half this in their short direction.
Spectra of lensed quasars and close quasar pairs can be used to
constrain the characteristic transverse sizes of Ly$\alpha$ absorbers.
On very small (few kpc) scales, absorbers are virtually identical in
both sight-lines (Smette et al. 1992; Dolan et al. 2000).  On
intermediate (tens of kpc) scales spectra are very similar but are
currently too low-resolution to conclusively constrain the absorber
properties across the sight-lines (e.g., Bechtold et al. 1994; Smette
et al. 1995).  On the largest scales, statistical analyses of the
probability of detecting an absorber in both sight-lines lead to
estimates of absorber `sizes' of several hundred kpc (e.g., Dinshaw et
al. 1995; Smette et al. 1995; Crotts \& Fang 1998; D'Odorico et
al. 1998).  These observations are slightly at odds with the MOND
predictions unless $g_0$ is near its upper limit, but some additional
considerations must be kept in mind:

\begin{enumerate}
\item{Except on very small scales, the observations do not currently
rigorously distinguish between correlated absorbers and a single
absorber spanning two sight-lines.  So we cannot directly compare the
observationally deduced sizes (several hundred kpc) to the Jeans
length of absorbers of the observed column density
(Eqs.~\ref{eq-absz} and~\ref{eq-absz2} ).}
\item{Strictly speaking, the relations given in Eqs.~\ref{eq-absz}
and~\ref{eq-absz2} connect observed column density to radial dimension
whereas the observations probe the transverse extent (or correlation).
However, for an `isolated' absorbing region these dimensions are
unlikely to be very different because absorbers should vary
transversely on a scale comparable to the local Jeans length, which is
in turn comparable to the radial extent over which the absorption
occurs.  In the case of a strong external field, the absorbers should
be elongated in the field direction by a factor of two.}
\item{The properties of absorbers derived here assume that the
absorbers are not far from local dynamical equilibrium, which would
{\em not} be true for underdense and/or very large ($\gg 100\,$kpc)
absorbers with dynamical times exceeding the Hubble time. For isolated
absorbers, it can be shown that regions of this size could not give
rise to sufficient absorption unless their dynamical time {\em
is} shorter than the Hubble time.  To see this, note that requiring
$t_{\rm dyn} > t_H$ for a fixed $L$ gives an upper limit on density
via Eq.~\ref{eq-tdyn}, and this in turn gives an upper limit to column
density of
$$
N_{HI} < 4.4\times10^{11}{\rm\,cm^{-2}}L_{100}^3
\left({t_H\over
10^{17}\,{\rm s}}\right)^{-8}{T_4^{-0.76}\over \Gamma_{12}},
$$
where $L_{\rm 100}=L/100\,{\rm kpc}$.  This means that isolated $\sim
100\,$kpc absorbers of column density $\sim 10^{14}{\rm\,cm^{-2}}$
cannot consistently be far from local dynamical equilibrium.
Absorbers in an external potential $g_0$ can be out-of-equilibrium if
their column density obeys
$$
N_{HI} < 10^{14}{\rm\,cm^{-2}}L_{100} \left({t_H\over
10^{17}\,{\rm s}}\right)^{-4}\left({a_0/g_0\over25}\right)^{-2}{T_4^{-0.76}\over
\Gamma_{12}}.
$$
Thus absorbers in a strong external field may be only marginally in
equilibrium.}
\end{enumerate}

	In summary, we find that if the dynamics of Ly$\alpha$
absorbers in MOND are dominated by their self-gravity, then they are
significantly smaller than observations indicate.  If, however, the
clouds are immersed in a (constant) acceleration field of magnitude
$g_0$, then their sizes can be substantially larger, and the basic
properties of Ly$\alpha$ absorbers approach those predicted by standard
gravity (with CDM) as $g_0 \rightarrow (\Omega_{\rm gas}/\Omega_{\rm
dm})a_0 \approx 0.16a_0$, where $\Omega_{\rm gas}$ and $\Omega_{\rm
dm}$ are the density parameters in absorbing gas and dark matter in
the CDM model.  The importance of external fields in low-acceleration
systems is both a methodological barrier and a saving grace of
MOND.  The accurate description of such systems requires a
(necessarily ab-initio) calculation of the large-scale density field
surrounding them, which is in turn impossible to perform rigorously
without a cosmological formulation of MOND which treats its conceptual
problems (in particular the issue of which accelerations, and with
respect to what, should be `counted').  On the other hand, external
fields can ensure that in the limit of extremely low accelerations the
properties of isolated systems in MOND will not deviate wildly from
their Newtonian counterparts.

\section{MOND and clusters}

	Like Ly$\alpha$ systems, clusters of galaxies are
gas-dominated (in the absence of dark matter), well observed (this
time via X-ray measurements), should be in local hydrostatic
equilibrium, and are in the MOND acceleration regime.  They also have
internal accelerations larger than expected ambient acceleration
fields (and comparable to those near galaxies).  Thus strong
predictions regarding their structure in MOND can be made using
relatively simple arguments, as follows.

	For a gaseous system of mass $M$ in hydrostatic equilibrium,
$M$ is close to the Jeans mass $M_J$.  For pure gas with MONDian
gravity,
\begin{eqnarray}
\label{eq-mondmt}
M_J &\equiv& \rho L_J^3 = {1\over a_0 G}\left({\gamma k T\over \mu m_p}\right)^2 \\ \nonumber
&\approx& 4.6\times10^{12}\msol\left({kT\over {\rm keV}}\right)^2,
\end{eqnarray}
using Eq.~\ref{eq-mondjl}.
For comparison, the
observed mass-temperature relation of clusters found by
Mohr, Mathiesen, \& Evrard (1999) is
\begin{equation}
M^{500}_{\rm ICM} 
	\approx 4.3\times10^{12}\msol\left({kT\over {\rm keV}}\right)^{1.98},
\label{eq-obsmt}
\end{equation}
where $M^{500}_{\rm ICM}$ is the enclosed mass of the X-ray gas at the
radius within which its density is approximately 500 times the cosmic
mean.  While only an order-of-magnitude estimate, Eq.~\ref{eq-mondmt}
would at first seem to be a remarkable success of MOND, since the CDM
picture has difficulty explaining the observed mass-temperature
relation in detail (see, e.g., Mohr et al. 1999; Finoguenov, Reiprich
\& B\"ohringer 2001 and references therein).  The fact that MOND can
roughly account (to within a factor of two) for the mass discrepancy
in clusters has been pointed out by Sanders (1999).  But a hint of
trouble is suggested by the fact that Eq.~\ref{eq-obsmt} applies for a
particular radius, whereas for isothermal clusters Eq.~\ref{eq-mondmt}
does not.  A closer look reveals that serious problems arise when the
temperature profile $T(r)$ predicted by MOND for a given density
profile $\rho(r)$ is compared to observations.

Consider a spherical system such as a cluster in hydrostatic equilibrium.
Then the temperature and density obey
\begin{equation}{1\over \rho}{dP \over dr}=-a,\end{equation}
 where $a$ is the magnitude of the radial gravitational
acceleration, and $P=(kT/\mu m_p)\rho$ is the pressure.
For $a \ll a_0$ in MOND, this can be rewritten as
\begin{equation}
{d\log \rho\over d\log r}+{d\log T\over d\log r}=
	-{\mu m_p\over kT}[a_0 G M(r)]^{1/2},
\label{eq-he}
\end{equation}
where $M(r)$ is the enclosed mass.  This equation immediately implies
that if $\rho(r)$ and $T(r)$ are power laws, then $T(r) \propto
\sqrt{M(r)}$, as per the Jeans argument.  More generally, an
increasing $M(r)$ implies that $T(r)$ increases as long as the sum of $\alpha_\rho
\equiv d\log\rho/d\log r$ and $\alpha_T \equiv d\log T/d\log r$ changes
more slowly than $M(r)^{1/2}$.  Thus isothermal density profiles
in MOND tend to have a core which contains most of the
mass (and in which the logarithmic derivatives change relatively
quickly), then fall off more quickly than $r^{-3}$ (see Milgom 1984).
We will show that this is {\em not} the case for the X-ray emitting
gas that dominates the mass in observed clusters, yet clusters are
observed to be nearly isothermal outside of their central regions.

\begin{figure*}
{\vbox{ \centerline{ \epsfig{file=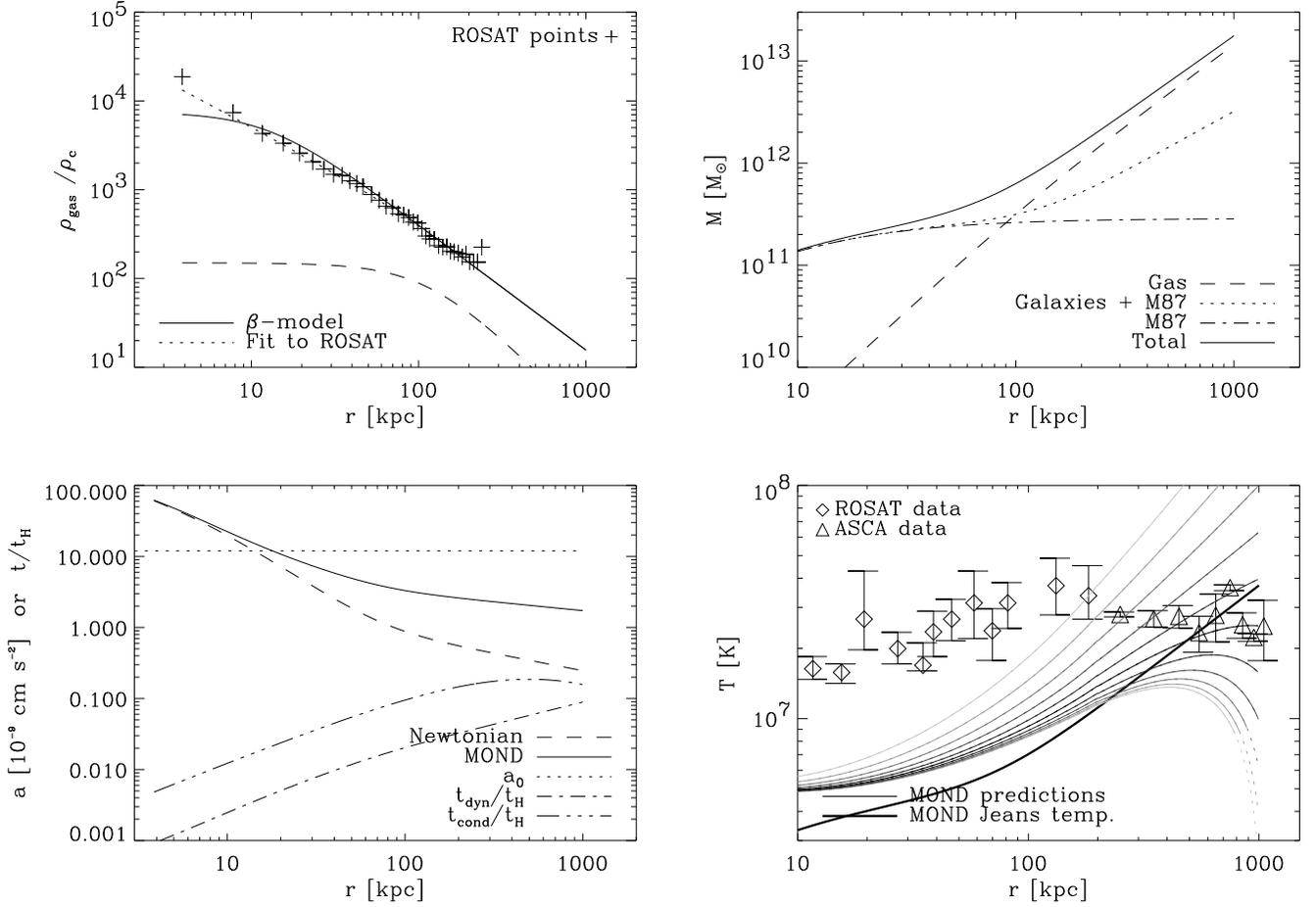,width=18.0truecm}}
\figcaption[]{ \footnotesize Predicted MOND temperature profile for
the Virgo cluster.  {\bf Top left:} X-ray gas density profile in units
of the critical density, from Nulsen \& B\"ohringer (1995) (pluses).
The dotted line is a Hernquist model fit at (and used here at) $r \la
200\,$kpc by Giraud (1999); the solid line is a $\beta$-model, fit by
Schindler et al. (1999) at large radii (solid line) and used here at
$r > 200\,$kpc.  {\bf Top right:} integrated mass in gas and galaxies.
The mass of M87 is from Giraud (1999) with $M/L_B = 8$; we also add a
component representing galaxies from Schindler et al. (1999). {\bf
Bottom left:} The Newtonian and MONDian acceleration (in units of
$10^{-9}\,{\rm cm\,s^{-2}}$) at each radius, showing that the cluster
is deep in the MOND regime for $r \gg 20\,$kpc. Also plotted are the
MOND dynamical time $t_{\rm dyn}$ and the conduction timescale $t_{\rm
cond}$, both in units of the $z=0$ Hubble time. {\bf Bottom right: }
Predicted temperature profiles, starting at 1\,Mpc with temperatures
between 0.1 and 10 times the (ASCA) observed temperature there.  ROSAT
(Nulsen \& B\"ohringer 1995, deprojected) and ASCA (Shibata et
al. 2001, projected) temperature profiles with 1$\sigma$ error bars
are shown for comparison.  The dark, solid line is the MOND `Jeans
temperature', $T_J \equiv (\mu m_p/k)(a_0GM)^{1/2}$.
\label{fig-cluster}}}
\vspace*{0.5cm}
}
\end{figure*}

\begin{figure*}
{\vbox{ \centerline{ \epsfig{file=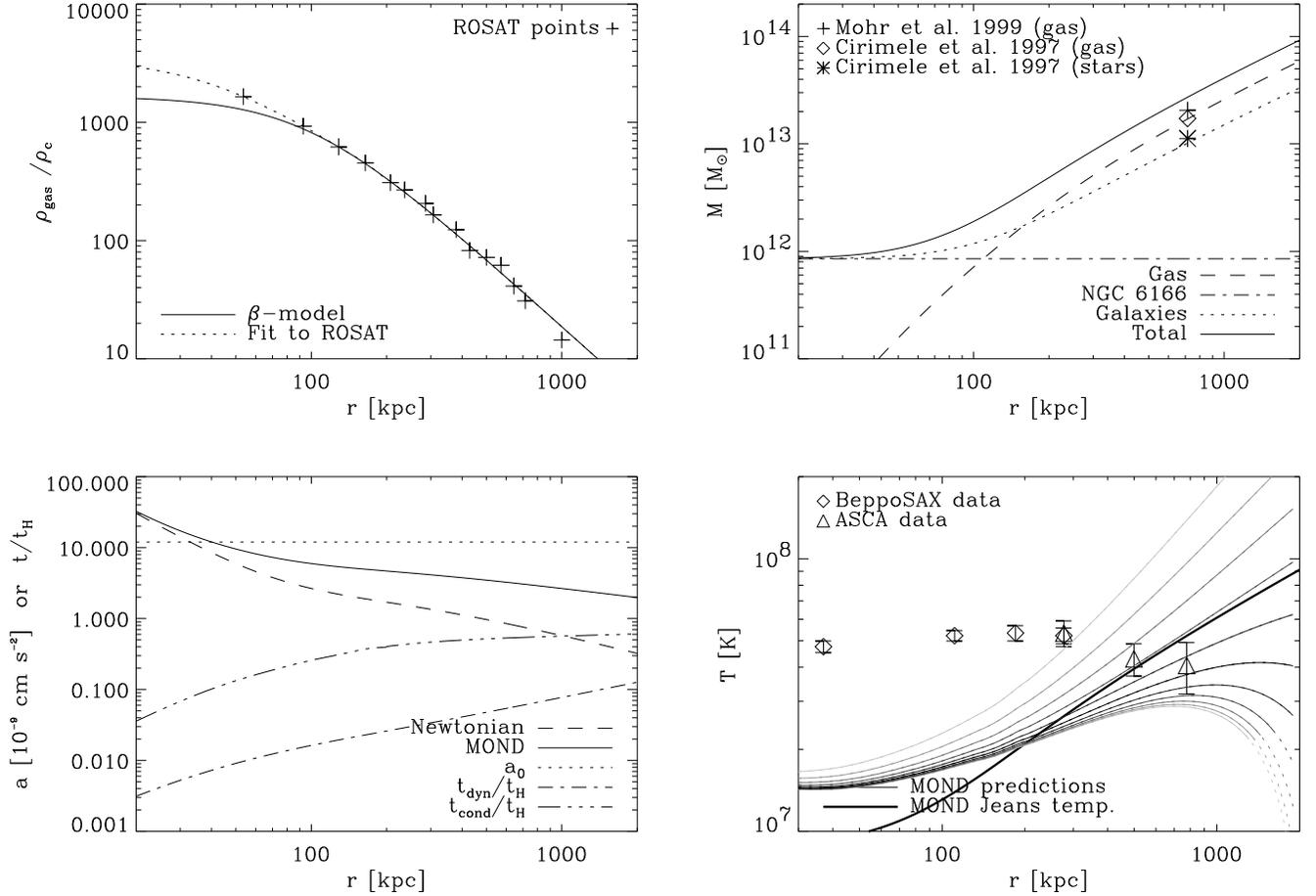,width=18.0truecm}}
\figcaption[]{ \footnotesize Predicted MOND temperature profile for
Abell 2199.  {\bf Top left:} X-ray gas density profile in units of the
critical density, from Siddiqui, Stewart, \& Johnstone (1998) (pluses;
dotted line is a spline-fit), with the $\beta-$profile of Markevitch
et al. 1999 (solid line) fit at large radii and used here at $r >
200\,$kpc.  Adjusted to $h=0.7$. {\bf Top right:} integrated mass in
gas and galaxies.  The central galaxy NGC 6166 is taken
(conservatively) to be a point mass with $M_V = -23.47$ (Gebhardt et
al. 1996) and $M/L_V=8$; we also add a component representing galaxies
proportional to the gas mass as per Cirimele et al. (1997) with
amplitude chosen so to give total mass fraction in stars at
$1h_{50}^{-1}\,$Mpc equal to theirs.  {\bf Bottom left:} The Newtonian
and MONDian acceleration at each radius, showing that the cluster is
deep in the MOND regime for $r \gg 30\,$kpc.  Also plotted are the
MOND dynamical time $t_{\rm dyn}$ and the conduction timescale $t_{\rm
cond}$, both in units of the $z=0$ Hubble time. {\bf Bottom right: }
Predicted {\em projected} temperature profiles, starting at 2\,Mpc
with temperatures between 0.1 and 10 times the (ASCA) observed
temperature there.  BeppoSAX (Irwin \& Bregman 2000) and ASCA
(Markevitch et al. 1999) temperature profiles with 2$\sigma$ errors
are shown for comparison. The dark, solid line is the MOND `Jeans
temperature', $T_J \equiv (\mu m_p/k)(a_0GM)^{1/2}$.
\label{fig-cluster2}}}
\vspace*{0.5cm}
}
\end{figure*}

\begin{figure*}
{\vbox{ \centerline{ \epsfig{file=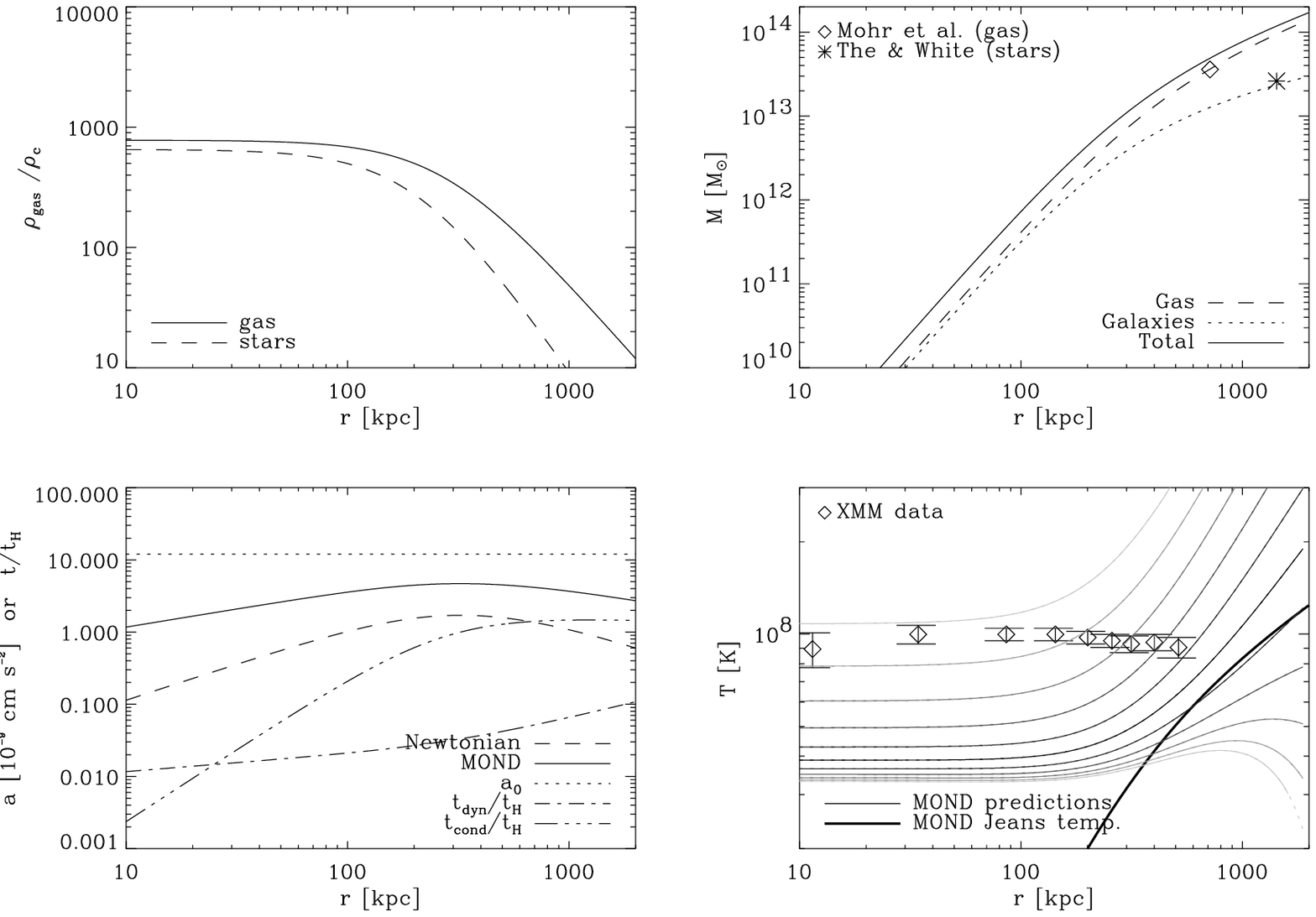,width=18.0truecm}}
\figcaption[]{ \footnotesize Predicted MOND temperature profile for
Coma.  {\bf Top left:} X-ray gas density profile (solid), using the
$\beta-$model fit of Mohr et al. (1999), adjusted to $h=0.7$. The
dashed line is the galaxy stellar mass density, using $M/L_V=8$ and
luminosity density from The \& White (1988).  {\bf Top right:}
integrated mass in gas and stars.  {\bf Bottom left:} The Newtonian
and MONDian acceleration at each radius, showing that the cluster in
the MOND regime for all radii.  Also plotted are the MOND dynamical
time $t_{\rm dyn}$ and the conduction timescale $t_{\rm cond}$, both
in units of the $z=0$ Hubble time. {\bf Bottom right: } Predicted {\em
projected} temperature profiles, starting at 2\,Mpc with temperatures
between 0.1 and 10 times the observed temperature there.  The XMM
temperature profile (Arnaud et al. 2001a) with 2$\sigma$ errors is
shown for comparison. The dark, solid line is the MOND `Jeans
temperature', $T_J \equiv (\mu m_p/k)(a_0GM)^{1/2}$.
\label{fig-cluster3}}}
\vspace*{0.5cm}
}
\end{figure*}

This constitutes a grave challenge to MOND which can be demonstrated
using the specific cases of the Virgo, Abell 2199, and Coma clusters,
for which gas density, stellar mass, and temperature profiles are
available in the literature.  The analysis is shown in
Figs.~\ref{fig-cluster}, ~\ref{fig-cluster2} and ~\ref{fig-cluster3}.
The first panel of Fig.~\ref{fig-cluster} shows the cluster gas
density deprojected from a ROSAT X-ray emission profile by Nulsen \&
B\"ohringer (1995), normalized to the critical density, assuming a
distance to M87 of $16\,$Mpc.  The density can be extended to $r >
200\,$kpc using the $\beta-$model fit by Schindler et al. (1999) at
large radii. The second (top-right) panel shows the integrated gas
mass, as well as integrated stellar mass from Schindler et al. (1999),
with a contribution for M87 with $M/L_B = 8$ included from Giraud
(1999).  As discussed below, the details of these assumptions matter
very little.  The bottom-left panel shows the Newtonian acceleration
at $r$, which is below the MOND parameter for $r \ga 20\,$kpc; the
MONDified acceleration\footnote{We use the same interpolation formula
in the transition region as Milgrom 1983b and Begeman, Broeils, \&
Sanders(1991), i.e.  $a\mu(a/a_0)=a_N$, where
$\mu(x)=x/(1+x^2)^{1/2}$.} is also plotted, along with the dynamical
time $\sqrt{l/a}$ in units of the $z=0$ Hubble time.  Given this
information, Eq.~\ref{eq-he} can be used to predict $T(r)$ given a
starting $T(r_0)$.  The lower-right panel shows this prediction,
integrating inward\footnote{The integration can also be performed
outward, matching the observed temperature at small radii, with
essentially the same results.} starting at $r_0 \approx 1\,$Mpc, with
$T(r_0)$ taking values between 1/10th and ten times the measured ASCA
temperature there.  If MOND were correct, one of these profiles should
roughly match the observed temperatures, but none of them do.
Figures~\ref{fig-cluster2} and ~\ref{fig-cluster3} shows the same
analysis for Abell 2199 and Coma, somewhat richer and more relaxed
clusters.  The results are quite similar.

	We have verified that the results are robust to reasonable
changes in the distances to the clusters, the mean molecular
weight, and the MOND interpolation formula used.  We have also
experimented with different profiles and normalizations for the
stellar mass distribution; these do not significantly affect the
results unless the stellar mass is so large ($M/L \gg 20$) as to imply 
the presence of dark matter (dark matter is discussed below).
Significantly larger values of the MOND constant
($a_0 \ga 5\times 10^{-8}\,{\rm cm\,s^{-1}}$) help improve the fit
because (as demonstrated by the agreement of Eqs.~\ref{eq-mondmt}
and~\ref{eq-obsmt}), the MOND `Jeans Temperature' $T_J^{\rm MOND}
\equiv (\mu m_p/k)(a_0GM)^{1/2}$ roughly agrees with observed cluster
temperatures {\em at large radii}.  Increasing $a_0$ moves this
agreement to intermediate radii, but still cannot yield a reasonable
fit of the entire profile (and would be incompatible with the value
required by galaxy rotation curves and make Ly$\alpha$ absorbers even
smaller).


	As alluded to above, the difficulty in accounting for the
cluster data in MOND -- even when $a_0$ is allowed to vary -- can be
understood in more general terms using an inequality derived from
Eq.~\ref{eq-he} which applies to any range $[r_1,r_2]$ over which
the temperature is non-increasing:
\begin{equation}
{\alpha_\rho(r_2)+\alpha_T(r_2)\over \alpha_\rho(r_1)+\alpha_T(r_1)} \ge
\left[{M(r_2)\over M(r_1)}\right]^{1/2}.
\end{equation}
Convective stability requires that $|\alpha_T| <
{2\over3}|\alpha_\rho|$ lest entropy gradients be erased on a sound
crossing time (Sarazin 1988, p. 165), giving
\begin{equation}
{M(r_2)\over M(r_1)} \le \left[{5\alpha_\rho(r_2) \over 3\alpha_\rho(r_1)}\right]^2.
\end{equation}
In Virgo, for example, $M(r)$ increases by a factor of
$\approx 25$ between 100 and 1000\,kpc, where the gas is observed to
be roughly isothermal. But then $\alpha_\rho(r=100\,{\rm kpc})\approx
-1.3$ requires (in MOND) that $\alpha_\rho(r=1000\,{\rm kpc}) \approx
-3.9$, while $\alpha_\rho \approx -1.5$ is observed.  Both power-law indices
and total baryonic mass are very well constrained quantities, so this is
a serious violation.  Cluster gas density profiles are generally
well-fit at large radii by $\beta$-models of form
\begin{equation}
\rho(r)=\rho_0\left[1+\left({r\over r_0}\right)^2\right]^{3\beta/2}.
\end{equation}
In this case $\alpha(r)/\alpha(r_0)=2/[1+(r_0/r)^2] < 2$.  Therefore
if $M(r)/M(r_0) > (10/3)^2 \approx 11$ at any radius within which the cluster
has a non-rising temperature, and within which the density profile is
well fit by a $\beta-$model, then MOND is violated.  If $T(r)$ is
exactly constant, the constraint is stronger and $M(r)/M(r_0) > 4$
violates MOND.

Since the relevant properties of Virgo, A2199 and Coma (isothermal or
radially declining temperatures, increasing $M(r)$, and slowly
changing $d\log\rho/d\log r$ as in the $\beta$-model) seem generic in
clusters at large radii (Neumann \& Arnaud 1999; Finoguenov, David, \&
Ponman 2000; Irwin, Bregman, \& Evrard 1999), it is very hard to see
how to reconcile MOND with the observations.  A few possibilities
which do {\em not} seem able to satisfactorily effect this
reconciliation are:

\bigskip
{\em Clusters are not in hydrostatic equilibrium:} The $z=0$ Hubble
time greatly exceeds the dynamical time (see bottom-left panel of each
figure) and the sound-crossing time, inside $\sim 1$\,Mpc, so
hydrostatic equilibrium should hold within that radius (see Sarazin
1988 for some discussion).  Moreover, simulations (albeit in the CDM
picture) show that hydrostatic equilibrium and spherical symmetry are
good assumptions in inferring the gravitational force given observed
density and temperature profiles (Schindler 1996).  Finally, we note
that the observed magnetic fields in clusters do not appear
sufficiently strong to significantly affect the force balance or the
inferred cluster bulk properties (e.g., Goncalves \& Friaca 1999;
Dolag, Evrard, \& Bartelmann 2001).

{\em Measured temperatures are incorrect:} while temperature
determinations do have significant errors (especially in ROSAT data),
the predicted MOND temperature profile disagrees by many $\sigma$ from
the ASCA measurements in Virgo, from ASCA and BeppoSAX measurements in
A2199, and from XMM measurements in Coma.  More generally, there is no
indication in observations using ASCA (e.g., Markevitch et al. 1998;
White 2000; Finoguenov, Arnaud \& David 2001), BeppoSAX (Irwin \&
Bregman 2000), or XMM (Arnaud et al. 2001a,b) that clusters have
steeply rising temperatures at large radii.

{\em Efficient conduction causes clusters to be isothermal:}
Figure~\ref{fig-cluster} shows gives the conduction timescale
(Sarazin 1988)
\begin{equation}
t_{\rm cond} \approx {5n_er^2k\over 2\kappa(T)},
\end{equation}
where 
\begin{equation}
\kappa(T) = 2\times 10^{11}\left({kT\over{\rm keV}}\right)^{5/2}{\rm
erg\,s^{-1}\,K^{-1}}.
\end{equation}
This is a lower limit, because it neglects magnetic fields, which
could increase the conduction time by a factor of between several and
several thousand (see, e.g., Rosner \& Tucker 1989 and Chandran \&
Cowley 1998, respectively).  If $t_{\rm cond} \la t_{\rm hub}$ for
some range in radius, we would expect to see a nearly isothermal
temperature profile there.  However (and regardless of whether or not
$t_{\rm cond} \la t_{\rm dyn}$), the cluster density profile would
respond to this conduction by readjusting to restore hydrostatic
equilibrium on a dynamical time.  But then (by an inverted version of
the argument given above), MOND would predict a density profile with
quickly-varying logarithmic derivatives, contradicting the observed
density profile.  In other words, Eq.~\ref{eq-he} prescribes a 1-to-1
relation between $\rho(r)$, $M(r)$ and $T(r)$, demonstrated in
Figs.~\ref{fig-cluster} and~\ref{fig-cluster2}.  Since the predicted
MOND $T(r)$ is not isothermal, an isothermal $T(r)$ cannot
match the observed $\rho(r)$.

{\em Observed density profiles are incorrect:} It has been proposed
that a multiphase medium in clusters in which one component has a
small filling factor could lead to an overestimate of gas the density in
X-ray measurements (e.g., White \& Fabian 1995; Gunn \& Thomas 1996).
If this effect existed and were more severe at large radii, the
observationally inferred density profile could be flatter than the
true profile, which could have more quickly-varying $\alpha_\rho$.
However, studies of this effect find that the mass discrepancy is
likely to be relatively small ($\ll 50\%$) and, moreover, {\em less}
important at large radii (White \& Fabian 1995; White \& Gunn \&
Thomas 1996; Nagai, Sulkanen, \& Evrard 2000).  The comparison between
masses inferred from Sunyaev-Zel'dovich measurements ($\propto n_e$)
and X-ray emission measurements ($\propto n_e^2$) also disfavor large
corrections (Grego et al. 2001; Patel et al. 2000; Nagai et al. 2000).

Based on observations of an apparent excess of extreme-UV emission, it
has also been claimed that clusters may contain a large mass of warm
($\sim 10^6\,$K) gas (e.g., Lieu, Bonamente, \& Mittaz 2000;
Bonamente, Lieu, \& Mittaz 2001).  If this is the case, it would be
extremely favorable to the MOND hypothesis.  Note, however, that
another group, while finding EUV excess, finds a significantly smaller
intensity (e.g., Bowyer, Korpela, \& Bergh{\" o}fer 2001), and that
FUSE observations place tight limits on warm gas in Coma and Virgo
which are inconsistent with the proposed models (Dixon et al. 2001).

{\em There is dark matter in clusters:} One might argue that there is
a strongly concentrated component of baryonic dark matter in clusters
(Milgrom 1999; Sanders 1999), leading to nearly-isothermal
temperatures where that component dominates the gravitational mass.
Using our calculations we can estimate the required amount of dark
matter.  To do so, we have fit dark matter density profiles of various
(NFW, $\beta-$model, Hernquist model) parametric forms\footnote{ We
have also tried several more general forms with no significant effect
on the conclusions.}, while constraining the ratio $\xi_1$ of enclosed
dark mass to enclosed gas mas at 1\,Mpc.  By lowering this ratio we
can find, for each parametric form, the lowest value of $\xi_1$ for
which a plausible fit can be found. All three models yield fairly good
fits if $\xi_1$ is left free, and yield $\xi_1 \approx 4-5$ for Coma,
$\xi_1 \approx 2-2.5$ in A2199 and $\xi_1 \approx 1-2$ in Virgo.  When
$\xi_1$ is fixed and progressively lowered, we find minimal allowed
values of $\xi_1 \approx 3$ for Coma, $\xi_1 \approx 1.5$ for A2199
and $\xi_1 \approx 1$ for Virgo.\footnote{Note that we are able to fit
temperature profiles -- in standard gravity -- using NFW profiles with
$7 \la \xi_1 \la 9$ and concentration parameters between $\sim 3-4$
(for Coma) and $\sim 7-10$ (A2199 and Virgo).  The concentration for
Coma is perhaps somewhat low but the results seem otherwise
reasonable.}  The analogous ratios at 200\,kpc take approximate
minimal values of 11, 13 and 7.5 respectively.  The details of the
fits depend upon $h$, $a_0$, the MOND interpolation formula and (of
course) the parametric form chosen, but the general result that MOND
requires dark matter of at least one and up to several times the gas
mass within 1\,Mpc, and about ten times the gas mass within 200\,kpc,
is robust unless $h < 0.5$ or $a_0 > 2\times 10^{-8}\,{\rm
cm\,s^{-2}}$.

\bigskip
The issues we have addressed in this section have been considered
before by Gerbal et al. (1992, 1993) who computed $M(r)$ using
observed (but extrapolated) $\rho(r)$ and the assumption of
isothermality.  This yielded $M(r)$ somewhat larger than the observed
mass.  Milgrom (1993) criticized this approach and suggested a
different one similar to that employed here.  Using the Coma cluster,
The \& White (1988) tested whether they could generate a set of
$\rho(r)$, $M(r)$ and $T(r)$ in MOND that could match the observations
available at that time.  They succeeded, but only by using $h=0.5$ and
$a_0=2\times10^{-8}\,{\rm cm\,s^{-2}}$ (which are not now
observationally viable).  We have reproduced their calculations, but
find that even with their parameters, the now-available accurate
temperature profile of Coma cannot be fit.

	Thus it seems that while MOND can account for the
non-Keplerian form of galactic rotation curves (and -- somewhat
surprisingly -- for the mass-temperature relation in clusters), it
cannot account for cluster density and temperature profiles in detail.
The difference in success derives from the fact that -- roughly
speaking -- in MOND an asymptotically isothermal temperature profile
corresponds to a point mass, whereas clusters are extended yet still
in the MOND acceleration regime.  In CDM cosmology the isothermality
of both galaxies and clusters is explained by assuming that both are
embedded in an isothermal dark halo which dominates the mass.

\section{Discussion and Conclusions}

	MOND as an alternative to the dark matter hypothesis has
generally fared quite well when applied to galaxies.  However,
galaxies can only test MOND in the limited regime of accelerations
of (0.1-1)$a_0$ and physical scales of $< 100\,$kpc.  It is therefore
important to discover whether MOND's success extends to systems at
much lower accelerations and/or much greater physical scales. To this
end, we have predicted various properties of Ly$\alpha$ absorbers and
galaxy clusters in MOND, using hydrostatic equilibrium arguments.
Both types of systems are observationally well constrained, and well
within the MOND acceleration regime.

	We find that as compared to their properties in the CDM
picture, the Ly$\alpha$ absorbers in MOND have somewhat higher
characteristic density and smaller characteristic size for a given
column density, potentially in conflict with absorption studies of
quasar pairs and lenses.  The magnitude of the effect, however,
depends upon the (unknown) magnitude of the external acceleration
field in which they embedded, since MOND violates the strong
equivalence principle,\footnote{This effect, incidentally, severely
limits the possibility of testing the MOND formula over many decades
of acceleration.} and if the ambient acceleration field is large then
the predicted absorber properties can approach those observed.  When
combined with the observed differential number density of absorption
lines, the analysis yields (for any assumed external field) a
reasonable density of intergalactic gas (as does the CDM picture).  A
more accurate prediction of the detailed physical and statistical
properties of Ly$\alpha$ absorbers in MOND -- which match observations
in detail in the CDM case -- will probably require a (presently
infeasible) ab-initio calculation of the large-scale acceleration
field and the absorbers themselves.

We note also that the MOND Jeans mass depends only on temperature:
\begin{equation}
M_J \approx 3.4\times10^6\msol T_4^2,
\end{equation}
and gives a rather small Jeans mass in the IGM which may dramatically
alter galaxy formation in MOND (though we have not pursued that issue
here).

	Simple arguments can also predict the mass-temperature
relation in clusters, and the temperature profile of any single
cluster given its observed gas density and enclosed mass profile. The
MOND-predicted M-T relation is impressively close to the observed one,
though this success seems coincidental, as it is sensitive to the
radius at which the enclosed mass is measured (since in MOND the Jeans
mass depends only on the temperature and the clusters are roughly isothermal).  

	Stronger constraints can be derived using the observed mass,
temperature, and density profiles of clusters: given hydrostatic
equilibrium, MOND directly predicts the relation between the three
quantities. For the form of mass and gas density profiles generally
observed, MOND predicts rising temperature profiles.  In the specific
cases of the Virgo, Abell 2199, and Coma clusters, we have shown that
MOND's predictions strongly disagree with measurements from ASCA,
ROSAT, BeppoSAX and XMM, and we see no reasonable way to effect a
reconciliation without recourse to large amounts of (presumably
baryonic) dark matter of an unknown type.  For these clusters, the
mass of such dark matter must exceed the gas mass within $1\,$Mpc by a
factor of $\sim 1-3$, and by a factor of about ten within $100\,$kpc.
Moreover, we have argued that the discrepancy applies to clusters in
general.  This may be interpreted as a failure of MOND to describe
cluster dynamics in terms of their observed baryonic content, or as a
bold prediction (Sanders 1999) that we have so far observed only a
minority of their baryonic content. (Discovery of such dark matter
would also constitute a serious crisis for the CDM model.)

	In conclusion, we find that although MOND can explain the
rotation curves of galaxies in a simple and compelling way, it is less
effective in extended systems such as clusters and (perhaps)
intergalactic gas clouds, in which the visible mass cannot be
described as a gravitational point mass when the system is in the MOND
regime.  This implies that dark matter (or perhaps some different
modification of gravity) is required to accurately describe such
systems.

\acknowledgements

We thank Simon White, Stacy McGaugh, Moti Milgrom, David Spergel, Ned
Wright, and Neta Bahcall for helpful suggestions.  This work was
supported by a grant from the W.M. Keck foundation.  EQ is supported
by NASA through Chandra Fellowship PF9-10008, awarded by the Chandra
X--ray Center, which is operated by the Smithsonian Astrophysical
Observatory for NASA under contract NAS 8-39073.

\end{document}